\documentstyle[aps,preprint]{revtex}
\input psfig
\def\baselinestretch{1.25}
\def\be{\begin{equation}}
\def\ee{\end{equation}}

\def\k{\mbox{\bf k}}
\def\p{\mbox{\bf p}}

\begin{document}
\draft
\preprint{PURD-TH-96-08, OSU-TA-27/96, hep-ph/9610477}
\date{24 October 1996; revised 24 February 1997}
\title{Resonant decay of Bose condensates}
\author{S. Khlebnikov$^1$ and I. Tkachev$^{2,3}$}
\address{
${}^1$Department of Physics, Purdue University, West Lafayette, IN 47907 
\\
${}^2$Department of Physics, The Ohio State University, Columbus, OH 
43210\\ 
${}^3$Institute for Nuclear Research of the Academy of Sciences of Russia
\\Moscow 117312, Russia}

\maketitle

\begin{abstract}
We present results of fully non-linear calculations of decay of the inflaton 
interacting with another  scalar field $X$. Combining numerical results 
for cosmologically interesting range of resonance
parameter, $q \leq  10^6$, with analytical estimates, we extrapolate them to 
larger $q$. We find that scattering of $X$ fluctuations
off the Bose condensate is a very efficient mechanism limiting
growth of $X$ fluctuations. For a single-component $X$, the resulting
variance, at large $q$, is much smaller than that obtained in the Hartree
approximation.

\end{abstract}


\narrowtext
In recent years, we have come to realize that the post-inflationary 
universe had probably been a much livelier place than was previously thought.
In many models of inflation, the decay of the inflaton field is not a slow 
perturbative process but a rapid, explosive one. At the initial stage
of this rapid process, called preheating \cite{KLS}, fluctuations of Bose
fields coupled to the inflaton grow exponentially fast, 
which can be thought of as ``parametric resonance'' \cite{para}, 
and achieve large occupation numbers. 
At the second stage, called semiclassical thermalization \cite{us},
the resonance smears out, and the fields reach a slowly evolving 
turbulent state with smooth power spectra \cite{us,wide}. 

The explosive growth of Bose fields leads to very large variances
of these fields close to the end of the resonance stage. That could result
in several important effects taking place shortly after the end
of inflation. These include symmetry restoration, baryogenesis, and
SUSY breaking \cite{effects}. To find out if these effects had indeed 
occurred, one needs a good estimate of the maximal size of Bose fluctuations.
The semiclassical nature of processes involving states with large 
occupation numbers allows us to treat decay of the inflaton, and any Bose 
condensate in general, as a classical non-linear problem with random initial 
conditions for fluctuations \cite{us}. This classical problem can be 
analyzed numerically.

In this Letter we report results of fully non-linear calculations
for the most interesting case when the coupling of a massive inflaton 
$\phi$ to some other scalar field $X$ is relatively large. That means,
more precisely, that the system is in the regime of wide parametric resonance 
\cite{KLS}, characterized by a large value of the resonance parameter 
$q$, $q\gg 1$. We have studied both expanding and static universes, to cover 
both post-inflationary dynamics and decays of possible other Bose
condensates. Our objective was to obtain an estimate for the maximal size 
of $X$ fluctuations, importance of which we emphasized above. 

Our results are as follows. We have found that scattering of $X$
fluctuations off the Bose condensate of $\phi$, which knocks inflaton
quanta out of the condensate and into low-momentum modes, is very 
efficient in limiting the size of $X$ fluctuations for large values 
of $q$, such as required \cite{KLS,wide} to produce particles much
heavier than the inflaton. This scattering process involves the condensate 
of zero-momentum inflatons and, for that reason, is especially enhanced,
cf. Refs. \cite{us,Son2,AC}. 
Fluctuations of $X$ can reach larger values for smaller values of $q$. 
The suppression of the maximal size of $X$ fluctuations for large $q$ 
significantly restricts the possibility of GUT 
baryogenesis after inflation, as well as the types of 
phase transitions that could take place after preheating.

Our present results should be compared with those obtained in the 
Hartree approximation. We find that for a single-component field $X$ 
in flat space-time, the Hartree approximation is inadequate for all
$q\gg 1$. A similar conclusion was made in Ref. \cite{wide} for the 
conformally invariant case of massless 
inflaton interacting with a massless field $X$,
based on our simulations of the fully non-linear problem for that case.
The Hartree approximation, with its characteristic positive
feedback of $X$ on the inflaton decay \cite{KLS,wide},
may still apply when $X$ has sufficiently 
many components; it remains to see if this can happen for realistic 
sizes of GUT multiplets.

In the model with a massive inflaton (Model 1 of Ref. \cite{wide}), 
the full scalar potential is 
$V_{1}(\phi,X)=\frac{1}{2}m^2\phi^2+\frac{1}{2} g^2 \phi^2 X^2+
\frac{1}{2} M_X^2 X^2$.
For comparison, we will sometimes present results
also for the model with massless inflaton (Model 2), in which the 
potential is
$V_{2}(\phi,X)=\frac{1}{4}\lambda \phi^4+\frac{1}{2} g^2 \phi^2 X^2$.
Both fields ($\phi$ and $X$) have standard kinetic terms and are 
minimally coupled to gravity.
We will often use rescaled variables: for Model 1,
$\tau=m\eta$, where $\eta$ is the conformal time; 
$\mbox{\boldmath $\xi$}=m {\bf x}$;
$\varphi=\phi a(\tau)/\phi(0)$; $\chi=X a(\tau)/\phi(0)$.
For Model 2, one should replace $m$ with $\sqrt{\lambda} \phi(0)$ in these
rescalings. Here $a(\tau)$ is the scale factor of the universe, 
defined so that $a(0)=1$, and $\phi(0)$ is the 
value of the homogeneous inflaton field at the end of inflation.
The resonance parameter $q$ is $q\equiv g^2 \phi^2(0)/4 m^2$ for
Model 1, and $q\equiv g^2/4\lambda$ for Model 2.

In order for resonance 
to fully develop in expanding 
universe, the resonance parameter $q$ should exceed a certain minimum
value, $q_{\min}$, which depends on the mass $M_{X}$ of $X$ \cite{wide}.
In Model 1, for $M_X=0$, $q_{\min}\sim 10^4$; for $M_X=10 m$,
$q_{\min}\sim 10^8$. 
We have simulated this model for $q$ up to $q=10^4$ in flat space-time
and for a few $q$ ranging from  $10^{4}$ to $10^{6}$ in the expanding 
universe. We have developed an analytical approach, and 
we use analytical estimates to extrapolate our results to larger $q$, 
needed to produce heavier particles.

The full non-linear equations of motion for $\varphi$  and $\chi$, 
which follow 
from the action described above, are solved directly in configuration space
with initial conditions corresponding to conformal vacuum at the end 
of inflation for all modes with non-zero momenta.
The initial conditions for $\chi$ are given in Ref. \cite{wide}; those 
for $\delta \varphi$ are obtained similarly, along the lines of Ref. \cite{us}.
Classical fluctuations evolve from quantum 
ones, and, in cases we consider, their typical initial sizes 
are much smaller than the scale of non-linearity.
Hence, initial evolution is linear with respect to fluctuations.
During the initial linear stage, the equation of motion for Fourier 
components of $\chi$ can be approximated as
${\ddot \chi}_{\k} + \omega^2_k(\tau) \chi_{\k} = 0$, where
\begin{equation}
\omega^2_k(\tau) = m_\chi^2 a^2+k^2 - {\ddot a}/{a}
+4q\varphi_0^2(\tau)  \,\, ,
\label{ome}
\end{equation}
$m_{\chi}\equiv M_{X}/m$, and $\varphi_0$ is 
the zero-momentum mode of $\varphi$; $\dot{\varphi}_0(0)=0$.
By virtue of our rescaling, $\varphi_0(0)=1$. 

Computations were done on $128^3$ lattices, for a single-component $X$.
For the case of expanding universe, we used $m^2=10^{-12} M^2_{\rm Pl}$.
Energy non-conservation in flat space-time typically was less then $10^{-3}$;
in expanding universe in linear regime our calculations closely reproduced
calculations in the Hartree approximation, which have much better accuracy.

\begin{figure}
\psfig{file=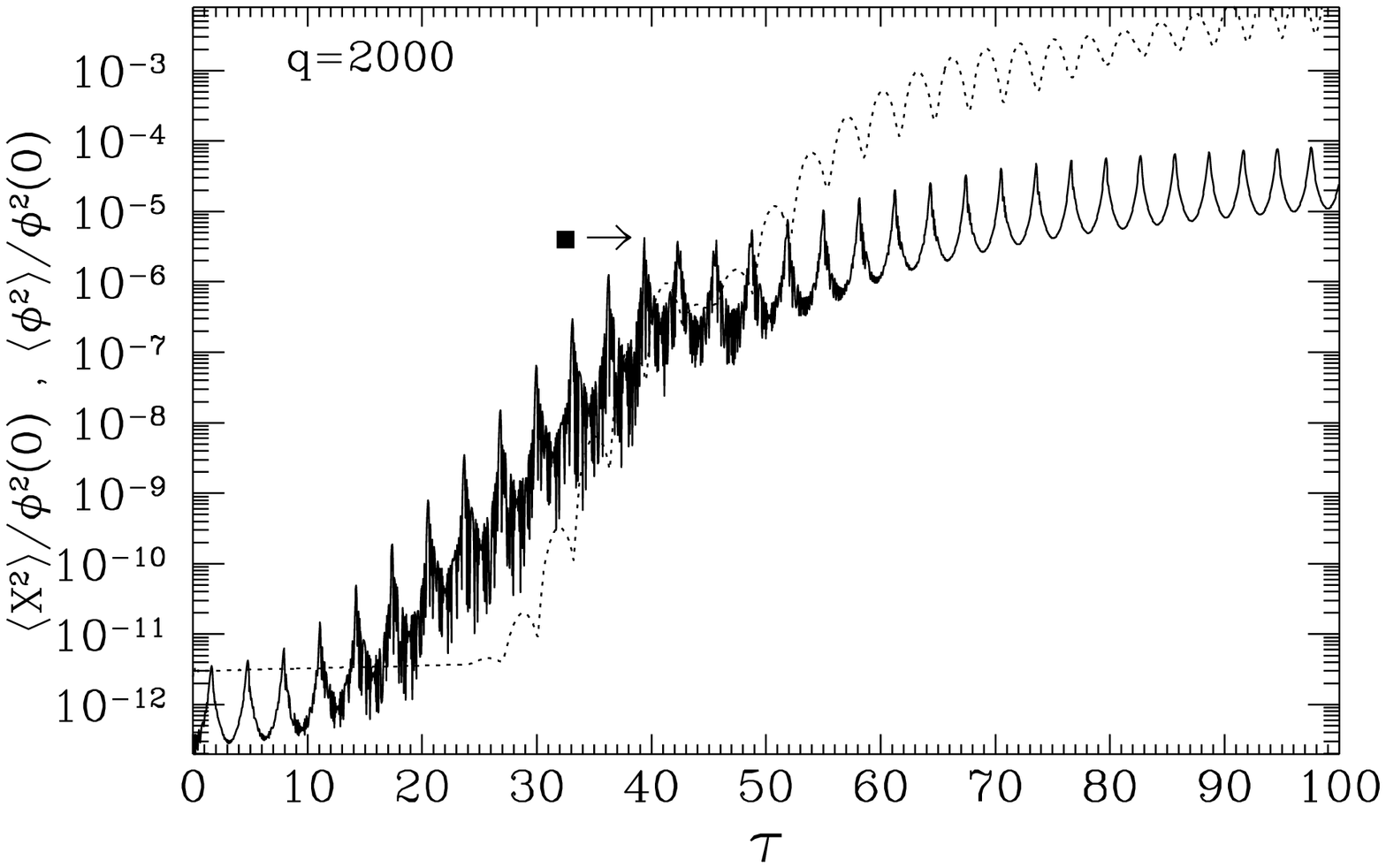,height=3.6in,width=5.6in}
\caption{Variances of the fields $X$ (solid curve) and $\phi$ (dotted 
curve) in Model 1 in flat space-time.
The filled square marks the spike value of $\langle X^2 \rangle$ at the 
end of resonance.} 
\label{fig:Fig1}
\end{figure}

Let us first consider the case without expansion of the universe. 
In the formulas above, one substitutes $a(\tau) =1$.
The variances of the fields $\chi$ and $\varphi$ 
in Model 1 at $q=2000$ as functions of time are shown in 
Fig. \ref{fig:Fig1}. The exponential growth of the variance 
$\langle \chi^2\rangle$ 
(the angular brackets denote averaging over space or, for space-independent
quantities, equivalently, over realizations of random initial conditions) 
at early times is a parametric resonance, which in the present 
case ends at $\tau \approx 40$. This marks the end of the linear stage. 

At large $q$, fluctuations of $X$ are produced at resonance stage
only during 
short intervals of time near moments when $\varphi_0$ passes through 
zero \cite{KLS,FKYY}. These are intervals in which the adiabatic (WKB)
condition ${\dot \omega}_{k}/\omega_{k}^{2}\gg 1$ is broken for some
$k$. Notice a series of spikes in $\langle \chi^2\rangle$ at the same 
moments of time.  They are due to
modulation of $\omega_k$  by the oscillating $\varphi_0$. Indeed,
introduce analogs of occupation numbers $n_{k}$ via
$\langle \chi^2\rangle = \int d^3k P_{\chi}(k) 
\propto \int d^3k n_{k}(\tau)/\omega_k(\tau)$.
Even at the resonance stage,  change in $n_{k}$ during one
oscillation is much smaller than variation of $\omega_k$:
when instead of $ \langle \chi^2(\tau)\rangle$ we plot $\int d^3k 
n_{k}(\tau)$, the spikes are replaced by relatively small steps at times
when $\varphi_0=0$.

We have monitored the power spectra, $P(k)$, of $\varphi$ and $\chi$ in all 
our integrations. The strongest resonant momentum of $\chi$
is typically of 
order $q^{1/4}$; for some $q$, though, it can be close to zero.
Development of resonance peaks for $\chi$ is followed
by appearance of peaks for $\varphi$ due to rescattering,
cf. Ref. \cite{us}. Later, rescattering 
leads to a turbulent state, characterized by smooth power spectra.

\begin{figure}
\psfig{file=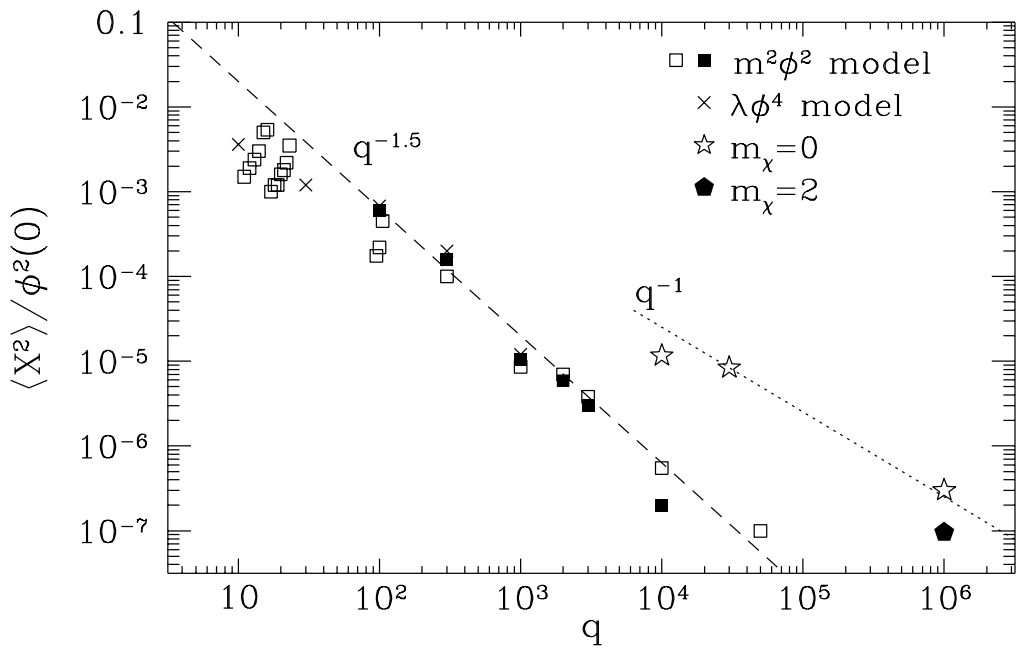,height=3.6in,width=5.6in}
\caption{Filled squares and crosses are the spike values, $\langle X^2 \rangle_{\rm s}$,
at the end of resonance obtained in fully non-linear simulations
of Model 1 (flat space-time) and Model 2; empty boxes are the spike
values at the first plateau in the Hartree approximation.
Stars correspond to $\langle X^2 \rangle$ at the moment when zero mode
decayed in Model 1 in expanding universe.} 
\label{fig:Fig4}
\end{figure}

Resonant growth stops when the Hartree correction to the mass of 
$\varphi$, due to $\langle \chi^2 \rangle$ (at spike, since this is when
fluctuations are produced) moves the system out of 
resonance. 
Because some modes grow faster than others,
the width of the resonance peak in the power spectrum of $\chi$ 
is much smaller than the full width (of order $q^{1/4}$) of the
instability band. Nevertheless, we find,
using analytical results of Ref. \cite{FKYY}, 
that the peak width still scales as $q^{1/4}$, up to a power of $\ln q$.
We then estimate that resonance ends when 
$\langle \chi^2\rangle_{\rm s} \sim q^{-3/2}$.
In Fig. \ref{fig:Fig4}, we have plotted results of our numerical
integrations for the spike value 
$\langle \chi^2\rangle_{\rm s}$ at the end of the resonance, as a function of 
$q$. At $q\agt 100$, these data are well fitted by 
$\langle \chi^2\rangle_{\rm s}\propto q^{-3/2}$. Figure \ref{fig:Fig4}
confirms that the termination of parametric resonance is a Hartree 
effect.

Parametric resonance is followed by a plateau 
(unless we consider an exceptional $q$, for which
the resonance peak was close to zero). There, the variances
of fluctuations do not grow, but an important restructuring of the 
power spectrum of $\chi$ takes place. 
The power spectrum of $\chi$ changes from being dominated 
by a resonance peak at some non-zero momentum to being dominated 
by a peak near zero.
The width of this new 
peak at $k \approx 0$ is of order one. 
When this peak becomes strong enough, the growth of 
variances resumes. The resumed growth 
(at $\tau\agt 50$ in Fig. \ref{fig:Fig1}) is quite rapid (compared to 
subsequent slow evolution) and is strongly affected by rescattering.  
This stage can be called semiclassical thermalization \cite{us}, or 
chaotization stage. Towards the end of it, the power 
spectra smoothen out; both power spectra are now dominated by 
momenta of order one.

An important effect seen in Fig. \ref{fig:Fig1} is the rapid growth
of fluctuations of the field $\varphi$. Indeed, at late times, they
are much larger than fluctuations of $\chi$. Fluctuations of $\varphi$
are produced by the scattering process in which $\chi$ fluctuations 
knock $\varphi$ out of the zero mode. 
If we neglected fluctuations of $\varphi$, i.e. the content of its
non-zero modes, the Hartree approximation in our model would 
be exact. As we will see, at large $q$ it can be far from being so.

At the end of the chaotization stage 
($\tau\approx 80$ in Fig.~\ref{fig:Fig1}), the variances in 
``valleys'' between the spikes of  $\langle X^{2} \rangle$
can be estimated directly 
from the classical equations of motion, solved in the limit 
$|\delta\varphi| \ll {\bar \varphi}$; here 
$\delta\varphi=\varphi-\varphi_{0}$, and ${\bar \varphi}$ is
the amplitude of  oscillations of $\varphi_{0}$. Using Green
functions, we obtain, for Fourier components of $\delta\varphi$ and $\chi$,
\begin{eqnarray}
\varphi_{\p}(\tau) & \approx & -\frac{4q}{\Omega_{p}}
\int_{0}^{\tau} d\tau' \sin[\Omega_{p}(\tau-\tau')]\varphi_{0} 
\int d^{3}k \chi_{\k}^{*} \chi_{\k+\p}  
\label{solphi} \\
\chi_{\k}(\tau)  & =  & \chi_{\k}^{(0)} + 8q\int_{0}^{\tau} dt' F_{k}(\tau,\tau') 
\varphi_{0} \int d^{3}p \varphi_{\p}^{*} \chi_{\k+\p}
\label{solchi}
\end{eqnarray}
where $\Omega_{p}=(p^{2}+1)^{1/2}$;
$\chi_{\k}^{(0)}(\tau) $ solves the Hartree equation, and $F_{k}(\tau,\tau')$ is 
the retarded Green function for it.
At times when $\chi$ grows, the main contribution to the
time integral in (\ref{solchi}) comes from $\tau'$ near $\tau$.
Using the random phase approximation, we then obtain 
$\langle |\varphi_{\p}|^{2}\rangle \sim 
q^{2}   {\bar \varphi}^{2} V^{-1}
\int d^{3} k \langle|\chi_{\k}|^{2}\rangle 
\langle|\chi_{\k+\p}|^{2}\rangle$, where $V$ is the total spatial volume.
The end of chaotization stage is the time, $\tau_{ch}$, when the 
second term in r.h.s. of (\ref{solchi}) becomes of the order of the first.
We find that that happens when, in terms of the physical fields,
\begin{equation}
\langle X^{2} \rangle_{v} \sim 
\frac{\phi^{3}(0)}{q^{3/2}{\bar \phi}(\tau_{ch})} \; ;
{}~~~~~~~\langle(\delta\phi)^{2}\rangle  \sim \phi^{2}(0)/q \; .
\label{varchi}
\end{equation}
where ${\bar \phi}$ is the amplitude of $\phi_{0}$.
In particular, $\langle X^2\rangle$ in``valleys" 
after rescattering, Eq. (\ref{varchi}), 
is of the order of $\langle X^2\rangle_{\rm s}$ at
the end of resonance. This is indeed seen in Fig. \ref{fig:Fig1}, as well 
as in all our other integrations in flat space-time.
Note that when variances of fields reach values (\ref{varchi}),
the analogs of occupation numbers, introduced as before, reach values
of order $1/g^{2}$ for {\em both} $X$ and $\phi$.

After chaotization, the maxima and minima of 
$\langle \chi^{2} \rangle$ evolve slowly, but the 
variance of $\varphi$ continues to grow rapidly for a while, see
Fig. \ref{fig:Fig1}. We interpret this as follows.
According to Eq. (\ref{solphi}), fluctuations of $\varphi$ are
driven by $\chi$. A periodic (or close to periodic) $\chi$ causes 
growth of $\delta\varphi$ via a usual, i.e. not parametric, 
resonance. This growth of $\delta\varphi$ will stop only when the
approximation leading to (\ref{solphi}) breaks down, that is
$\delta\phi\sim {\bar \phi}$.

\begin{figure}
\psfig{file=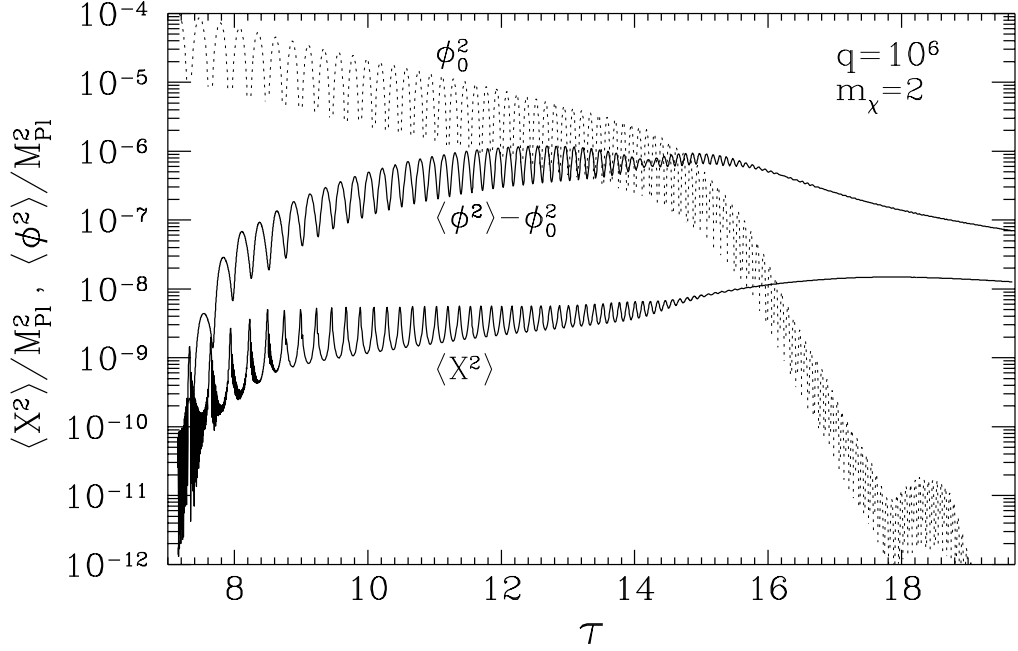,height=3.6in,width=5.6in}
\caption{Variances of fields $X$ and $\phi$,
together with the inflaton zero-momentum mode, in Model 1 
in expanding universe.} 
\label{fig:Fig5}
\end{figure}

\begin{figure}
\psfig{file=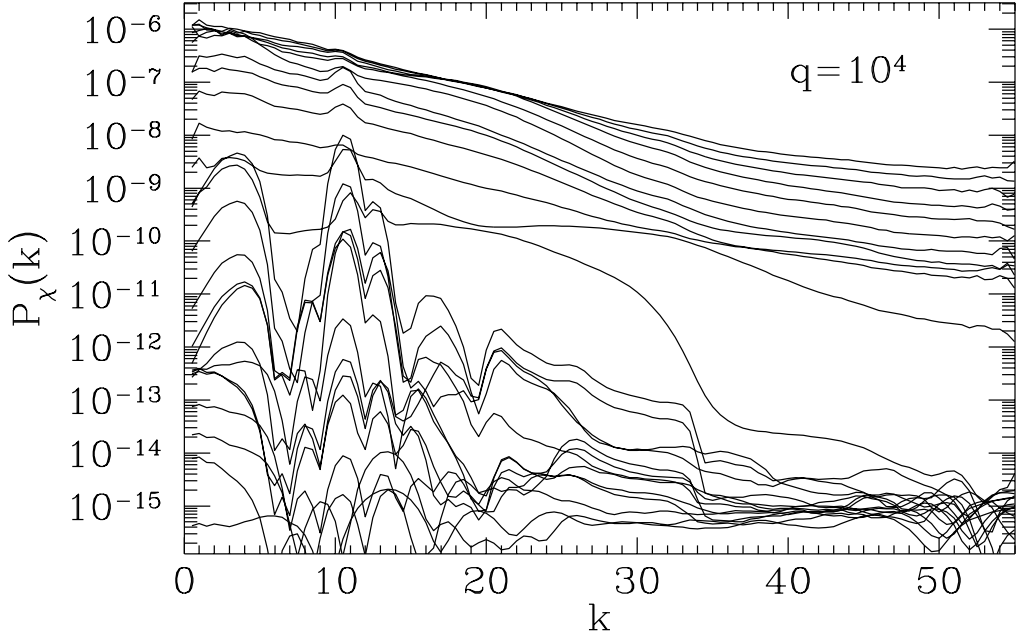,height=3.6in,width=5.6in}
\caption{Power spectrum of the field $X$
in Model 1 in expanding universe, output every period
at maxima of $\phi_0(\tau)$.}
\label{fig:Fig2}
\end{figure}

Let us now turn to Model 1 in expanding universe, see Figs. 
\ref{fig:Fig5}--\ref{fig:Fig2}. We used
$\phi(0)=0.28 M_{\rm Pl}$ \cite{wide}. For massless $X$, the evolution of 
the scale
factor was determined self-consistently, including
the influence of produced fluctuations in the Einstein equations;
for massive $X$, the universe was assumed matter dominated.

In expanding universe, particle creation acquires a
qualitatively new feature \cite{wide}: because of the time-dependence
of ${\bar \varphi}$, the resonance peak scans the entire 
instability band, see Fig. \ref{fig:Fig2}.  As our numerical integrations
confirm (see also \cite{wide}), in order for 
production of fluctuations to be efficient, variation of
the frequency of $\chi$, Eq. (\ref{ome}), need not be  
periodic. What is required is that every once in a while
the adiabatic condition breaks down. In this situation,
``non-adiabatic amplification" seems to be a better term for 
stimulated particle creation than ``parametric resonance''.

The variances of fields at the end of chaotization can be estimated in 
the same way as before, and we obtain again Eq. (\ref{varchi}). Note that
now ${\bar \phi}(\tau_{ch})\ll \phi(0)$, due to the redshift of the field.

Evolution of $\langle X^{2} \rangle$ after chaotization can be thought
of as a result of balance between creation and destruction of 
$X$ fluctuations. These processes are much faster than the expansion 
of the universe at this stage.
We may then assume that the system looses
memory of initial conditions and evolves as it would in flat 
space-time
but with the effective, slowly changing value of the resonance 
parameter,
$q_{\rm eff} (\tau) \equiv q {\bar \phi}^{2}(\tau)/\phi^{2}(0)$.
This assumption allows us to extend the estimate (\ref{varchi}) for 
$\langle X^{2}\rangle_{v}$ to times after chaotization by simply using 
the current value of ${\bar \phi}$ in it. In general, the assumption 
allows us to relate the scaling of variances with ${\bar \phi}$ to their 
scaling with $q$: 
$\langle X^2\rangle/{\bar \phi}^{2}\propto q_{\rm eff}^{-\alpha}(\tau)
\propto q^{-\alpha} {\bar \phi}^{-2\alpha}(\tau)$. 

Time dependence of variances is shown in Fig. \ref{fig:Fig5}
for the case $q=10^6$ and $m_\chi=2$. 
We can construct an analog of particle density, $n$, as
$n/m=[4q(\phi_0^2 +(\delta\phi)^2)/\phi^2(0) + m_{\chi}^2]^{1/2}
\langle X^2 \rangle$. As long as the term with $\phi_0$ is the leading 
term in the square brackets, $n\propto \langle X^2 \rangle_v {\bar \phi}$,
and according to our scaling argument 
[$\langle X^2 \rangle_v$ scales with $\alpha=3/2$, cf. Eq. (\ref{varchi})], 
$n$ after chaotization should be time-independent.
Our data confirms that.
Thus, fluctuations keep being produced at the same rate as they are 
diluted by the expansion. This can be called negative feedback amplification.
A striking feature of the data
is that,  after chaotization, $\langle X^{2}\rangle$ in spikes becomes 
essentially time-independent. According to our scaling argument, 
this means that $\langle X^{2}\rangle_{\rm s}$ scales as $1/q$. 
In Fig. \ref{fig:Fig4}, we show by stars the values of $\langle 
X^{2}\rangle$ in expanding universe, taken at times when $\phi_{0}$ 
decays (more precisely,
when ${\bar \phi}^{2}\approx \langle \delta\phi^{2}\rangle$) and
the ``spike'' and ``valley'' values of $\langle X^{2} \rangle$ coalesce.
(These values are within a factor of few from the maximal values 
$\langle X^{2}\rangle$ achieves in the expanding universe; the exact 
value is model dependent and it was largest, equal to 2.6 , for
$M_{X}=0$, $q=10^6$.) The scaling of these values with $q$ is well 
fitted by $1/q$.

We can now use the $1/q$ scaling of $\langle X^2\rangle_{\rm s}$
to extrapolate our data to
larger values of $q$. For example, for $q=10^{8}$, required to produce 
scalar leptoquarks with $M_{X}=10m$, we obtain 
$\langle X^2\rangle_{\rm s}\sim 10^{-10} M_{\rm Pl}^2$. 
Because $\langle X^2\rangle_{\rm s}$ stays on this level for a while,
despite the expansion, the time-integrated conversion of inflatons into 
$X$ fluctuations in this two-field model is in fact quite efficient.
For definite predictions for baryon asymmetry 
generated in decays of leptoquarks, however, one has to include other 
fields, which can have smaller $q$ and thus provide faster
alternative channels of inflaton decay.

If instead of a single-component $X$ we 
consider $X$ with $N$ real components, our estimate for
$\langle X^{2} \rangle_{v}$ at the end of chaotization becomes larger by a 
factor of $\sqrt{N}$. For realistic GUT values of $N$, this can 
increase the maximal value of $\langle X^{2}\rangle$ by an order of 
magnitude.
In comparison, in the Hartree approximation, the maximal value of 
$\langle X^2 \rangle/\phi^2(0)$ is of order 
$q^{-1/2} {\bar \phi}/\phi(0)$ \cite{KLS,wide}; for $q=10^{8}$ and
$M_{X}=10m$ we get
$\langle X^{2}\rangle_{\max}\sim 10^{-7} M_{\rm Pl}^{2}$. 

The relatively quick complete exponential decay of $\phi_{0}$, seen
in Fig. \ref{fig:Fig5}, 
is a distinctive feature of Model 1, as opposed to the 
conformally invariant Model 2. In the latter case, instead of
(\ref{varchi}), we obtain, at the end of chaotization,
$\langle(\delta\phi)^{2}\rangle  \sim {\bar \phi}^{2}/q$,
$\langle X^{2} \rangle_{v} \sim {\bar \phi}^{2}/q^{3/2}$, and
in the subsequent evolution $\langle X^{2}\rangle$ redshifts together
with ${\bar \phi}^{2}$. 

We thank L. Kofman, E. Kolb, A. Linde, M. Shaposhnikov for discussions 
and comments. The work of S.K. was supported
in part by the U.S. Department of Energy  under Grant DE-FG02-91ER40681 
(Task B), by the National Science Foundation under Grant PHY 95-01458, 
and by the Alfred P. Sloan Foundation. The work of I.T. was supported
by DOE Grant DE-AC02-76ER01545 at Ohio State.

As writing this paper was nearing completion, we received a preprint,
Ref. \cite{PR}, in which the method of Ref. \cite{us} was also applied
to two-field models. The authors of Ref. \cite{PR} consider only  flat
space-time and the conformally invariant case;
their results for these cases partially overlap with ours.


\def\baselinestretch{.8}

\end{document}